\documentclass[aip, pop, reprint, altaffilletter, floatfix]{revtex4-1}
\usepackage{graphicx}
\usepackage{float}
\usepackage{amsmath}
\usepackage{color}
\usepackage{multirow}

\begin{document}


\title{
Control and diagnosis of temperature, density, and uniformity in x-ray heated iron/magnesium samples for opacity measurements
} 



\author{T. Nagayama}
\affiliation{Sandia National Laboratories, Albuquerque, New Mexico 87185, USA}

\author{J. E. Bailey}
\affiliation{Sandia National Laboratories, Albuquerque, New Mexico 87185, USA}

\author{G. Loisel}
\affiliation{Sandia National Laboratories, Albuquerque, New Mexico 87185, USA}

\author{S. B. Hansen}
\affiliation{Sandia National Laboratories, Albuquerque, New Mexico 87185, USA}

\author{G. A. Rochau}
\affiliation{Sandia National Laboratories, Albuquerque, New Mexico 87185, USA}

\author{R. C. Mancini}
\affiliation{Physics Department, University of Nevada, Reno, Nevada 89557, USA}

\author{J. J. MacFarlane}
\affiliation{Prism Computational Sciences, Madison, Wisconsin 53703, USA}

\author{I. Golovkin}
\affiliation{Prism Computational Sciences, Madison, Wisconsin 53703, USA}


\date{\today}

\begin{abstract}

Experimental tests are in progress to evaluate the accuracy of the modeled iron opacity at solar interior conditions, in particular to better constrain the solar abundance problem [S. Basu and H.M. Antia, Physics Reports 457, 217 (2008)]. Here we describe measurements addressing three of the key requirements for reliable opacity experiments: control of sample conditions, independent sample condition diagnostics, and verification of sample condition uniformity. The opacity samples consist of iron/magnesium layers tamped by plastic. By changing the plastic thicknesses, we have controlled the iron plasma conditions to reach  i) $T_{e}$=167$\pm$3 eV and $n_{e}=\mathrm{\left(7.1\pm 1.5\right)\times10^{21}\, cm^{-3}}$, ii) $T_{e}$=170$\pm$2 eV and $n_{e}=\mathrm{\left(2.0\pm 0.2\right)\times10^{22}\, cm^{-3}}$, and iii) $T_{e}$=196$\pm$6 eV and $n_{e}=\mathrm{\left(3.8\pm 0.8\right)\times10^{22}\, cm^{-3}}$, which were measured by magnesium tracer K-shell spectroscopy. The opacity sample non-uniformity was directly measured by a separate experiment where Al is mixed into the side of the sample facing the radiation source and Mg into the other side. The iron condition was confirmed to be uniform within their measurement uncertainties by Al and Mg K-shell spectroscopy. The conditions are suitable for testing opacity calculations needed for modeling the solar interior, other stars, and high energy density plasmas.

\end{abstract}


\maketitle 

\section{Introduction}

Opacity quantifies photon absorption in matter and plays a crucial
role in many high energy density (HED) plasmas, including inertial
fusion plasmas and stellar interiors. Modeling opacity is especially
challenging when the plasma is composed of partially-ionized atoms
with many bound electrons. First, one must accurately compute atomic
data (e.g., energy level structures, photoionization cross-sections,
and oscillator strengths); then, atomic level populations have to
be computed for the given conditions including a sufficiently-complete
set of atomic levels; and finally, line shapes and spectral formation
must be taken into account in order to compute the frequency-dependent
opacity. The complexity of such models makes experimental tests critical.
However, the range of plasma conditions where opacity models are experimentally
validated is extremely limited because opacity measurement requirements
are also challenging. Opacity measurements at HED conditions, such
as those found in stellar interiors and inertial fusion ablators,
are particularly rare\cite{Bailey:2009hh}.

Without accurate opacity measurements, the systematic uncertainties
of modeled opacities are not known. This limits our understanding
of plasmas and sometimes complicates the interpretation of plasma
hydrostatic/hydrodynamic simulations when they disagree with measurements
or observations. Around 1980, it was reported that simulated Cepheid
variable stars could not reproduce the observed pulsations in certain
regimes\cite{Cox:1980ib}. In an attempt to resolve the discrepancies,
various modifications to the Cepheid models were tested, such as reducing
mass, increasing helium abundance, and introducing magnetic fields.
However, it turned out that the discrepancies were not caused by the
Cepheid models but mostly by inaccuracy of the calculated opacities
used in the models. Opacities re-computed using improved models were
significantly larger at Cepheid envelope conditions and resolved the
discrepancies\cite{Iglesias:1987be,Iglesias:1990fz,Rogers:1993ue}.
Calculating opacities is difficult, and this example illustrates the
possible consequences of their inaccuracy. 

More recently, calculated solar interior opacities, especially at
the base of the solar convection zone (CZ), are in question\cite{Bailey:2009hh}.
Standard solar models (i.e., hydrostatic models) were in excellent
agreement with helioseismic measurements until the solar metal abundances
were reduced in early 2000s \cite{Asplund:2005uv,Basu:2008fo,Asplund:2009eu,Basu:2013fo}.
This lowered the abundances of C, N, O, Ne, and Ar by 35-45\%, which
lowered the calculated solar mixture opacities accordingly. The downward
shift in the opacity altered the radiative heat transfer in the solar
models, affecting their agreement with helioseismic measurements,
especially at the CZ base (i.e., $R\sim0.713R_{SUN}$)\cite{Basu:2008fo}.
As with the Cepheid variable problem, the inaccuracy of calculated
opacity has been suggested as the origin of this CZ problem; a 10-30\%
increase in opacity would bring solar models and helioseismology into
agreement\cite{Basu:2008fo}. 

Until high-accuracy opacity measurements are conducted, opacity uncertainty
will continue to be a potential explanation for many HED and astrophysical
model-observation discrepancies. While a discrepancy often suggests
how much the mean opacity should be increased (or decreased), it does
not provide anything conclusive on calculated opacity accuracy. Are
the discrepancies really caused by opacity inaccuracy? Exactly how
are calculated opacities different from true opacities? Performing
benchmark experiments measuring opacity is crucial to answer these
questions and to guide potential opacity model refinements. Such refinements
would help to improve our understanding of the Sun, many other stars,
and a variety of other HED plasmas. 

Opacity measurement techniques have been developed and refined over
the past two decades\cite{Perry:1996hy,Davidson:1988hj,Perry:1991ez,Foster:1991cz,ChenaisPopovics:2001hr,Bailey:2003io,Renaudin:2006da,DaSilva:1992hd,Springer:1992hk,Winhart:1995el,Springer:1997iw,ChenaisPopovics:2000ct}.
Nevertheless, performing reliable experiments is demanding, and few
high quality measurements exist. There are many sources of difficulty\cite{Perry:1996hy,Bailey:2009hh}.
First, one has to tailor the sample to the target conditions while
also achieving sample uniformity, steady state, and local thermodynamic
equilibrium (LTE). When transmission is measured through significant
gradients, it is difficult to derive definitive conclusions on opacity.
Steady state and LTE are common assumptions in opacity models, and
thus have to be realized in the sample. To achieve LTE, the experiment
must either reach high densities or include a strong external radiation
field that is reasonably close to equilibrium with the sample plasma,
and these conditions must be maintained over a duration sufficient
to establish steady-state (i.e., $df_{i}/dt\sim0$ for each level
population $f_{i}$). Second, once the sample is tailored to the target
conditions, the sample transmission has to be accurately measured.
To benchmark the detailed atomic and plasma physics in modern opacity
models, it is necessary to measure the frequency-resolved transmission.
This requires a backlighter that is bright, to mitigate the sample
self-emission effects on the emergent transmission spectra\cite{Perry:1996hy,Bailey:2009hh},
and spectrally smooth, to avoid unnecessary complications due to instrumental
broadening\cite{Iglesias:2006dg}. Reliable frequency-resolved opacity
measurement also requires spectrometers with high resolving power
to observe detailed bound-bound line features. These spectrometers
must be free of any artifacts such as crystal defects and film scanning
flaws that could result in disturbing the data. Finally, the Fe plasma
conditions have to be independently measured to model the Fe opacity
and compare it with the measurement. 

In 2007, Fe opacity experiments were performed to constrain the CZ
problem discussed earlier, and Fe transmission spectra were successfully
measured at charge state distributions similar to Fe at the CZ base
for the first time\cite{Bailey:2007fq}. Fe was studied for two reasons.
First, Fe is one of the biggest opacity contributors at the CZ base.
Second, Fe at the CZ base has many bound electrons, making it more
difficult to model than many other elements in the solar matter. Uncertainty
of modeled Fe opacity may therefore be larger than the other elements.
The Fe samples were heated by the z-pinch dynamic hohlraum (ZPDH)
radiation source at the Sandia National Laboratories Z-machine. The
backlighter was provided by the stagnation of the ZPDH implosion and
was characterized as a smooth 314 eV blackbody radiator. This backlighter
was bright enough to mitigate sample self-emission at the measured
Fe conditions. The Fe reached an electron temperature, $T_{e}$, of
156 eV and an electron density, $n_{e}$, of 6.9$\times$10$^{21}\,\mathrm{cm^{-3}}$.
The conditions were independently measured by K-shell transmission
spectroscopy of Mg mixed with the sample. Fe L-shell transmission
spectra computed from detailed opacity models were shown to match
reasonably well with the measurements.

In order to benchmark Fe opacity models at the CZ base, more experiments
are needed. First, while the Fe transmission spectra were measured
at the charge state distribution similar to that of Fe at the CZ base,
the inferred $T_{e}$ and $n_{e}$ were significantly lower than the
conditions at the CZ base (i.e., 185 eV and 9$\times$10$^{22}\,\mathrm{cm^{-3}}$).
Thus, Fe opacity models need to be tested at higher $T_{e}$ and $n_{e}$
. This is important because additional effects become important as
$T_{e}$ and $n_{e}$ increase. In LTE, the same average ionization
can be achieved by increasing both $T_{e}$ and $n_{e}$. However,
since population within each charge state is distributed following
the Boltzmann relation, $p_{excited}/p_{ground}\propto\exp\left(-\Delta E/kT\right)$
where $\Delta E$ is the energy difference between the excited and
ground states, ions produced by higher $T_{e}$ and $n_{e}$ would
produce more population in the excited states. This makes the accuracy
of opacity calculations more sensitive to the accuracy of the atomic
data of excited levels and to the treatment of highly-excited levels
(doubly, triply, .., multiply excited levels). As density increases,
high density effects such as ionization potential depression and Stark
line broadening become more important. In particular, Stark line broadening
in L-shell line transitions has not been explored as extensively as
that of K-shell lines. Furthermore, even detailed modern opacity models
employ rather simple approximations similar to the formalism discussed
in Griem \textit{et al.}\cite{Griem:1968ba}. In order to disentangle
these physical details and benchmark opacity models, it is crucial
to control the plasma conditions and repeat Fe opacity experiments
at several different $T_{e}$ and $n_{e}$ while maintaining similar
charge state distributions. Such a collection of measurements would
promote investigations of how the effects of higher $T_{e}$ and $n_{e}$
gradually change the opacity and how well opacity models can predict
them. It is also important to make direct measurements of the sample
uniformity. In Bailey \textit{et al.}\cite{Bailey:2007fq}, the sample
uniformity assertion was supported by the use of volumetric heating
provided by ZPDH radiation and by the fact that the mixed Fe/Mg transmission
spectra were successfully modeled by a single $T_{e}$ and $n_{e}$.
However, it is preferable to experimentally quantify the level of
the sample non-uniformity. 

In this article, we describe measurements addressing three of the
key requirements for opacity experiments: i) control of the Fe conditions,
ii) independent plasma diagnostics, and iii) verification of the sample
condition uniformity. The Fe sample conditions can be controlled by
the thickness of the tamping plastic ($\mathrm{CH}$) as suggested
by hydrodynamic simulations \cite{Nash:2010cn}. Thus, we performed
Fe opacity experiments with three different $\mathrm{CH}$ tamper
thicknesses: i) 10 $\mu$m, ii) 35 $\mu$m, and iii) 68 $\mu$m. The
Fe conditions are independently measured by Mg K-shell spectroscopy
and found to be i) $T_{e}$=167$\pm3$ eV and $n_{e}=\mathrm{\left(7.1\pm1.5\right)\times10^{21}\, cm^{-3}}$,
ii) $T_{e}$=170$\pm2$ eV and $n_{e}=\mathrm{\left(2.0\pm0.2\right)\times10^{22}\, cm^{-3}}$,
and iii) $T_{e}$=196$\pm6$ eV and $n_{e}=\mathrm{\left(3.8\pm0.8\right)\times10^{22}\, cm^{-3}}$,
respectively. These conditions nearly reproduced the charge state
distribution of Fe at the CZ base, but with different levels of high
temperature and density effects. Also, one experiment was designed
and performed to directly investigate the sample non-uniformity. For
this particular experiment, Al was mixed into the side of the Fe sample
facing the radiation source and Mg in the opposite side. Fe conditions
in both sides are inferred by Al and Mg K-shell spectroscopy. We confirmed
that there is no measurable axial non-uniformity in the Fe sample.
The discussion of other requirements, such as steady-state and LTE,
as well as details of the transmission determination, the comparison
with Fe opacity models, and the implications to astrophysics, atomic
physics, and HED physics are beyond the scope of this article and
will be discussed elsewhere. Section \ref{sec:Z-pinch-dynamic-hohlraum}
discusses the experiments and data reduction. Section \ref{sec:K-shell-spectroscopy-for}
explains the Mg K-shell spectroscopy, how to model the spectra, and
how to find the optimal $T_{e}$ and $n_{e}$ from the measured Mg
spectra. Section \ref{sec:Fe-condition-analysis} summarizes Fe conditions
analyzed for experiments using different $\mathrm{CH}$ tamper thicknesses.
Section \ref{sec:Uniformity-measurement} summarizes the sample non-uniformity
investigation.

\section{\label{sec:Z-pinch-dynamic-hohlraum}Experiments and data reduction}

\begin{figure}
\includegraphics[width=8.5cm]{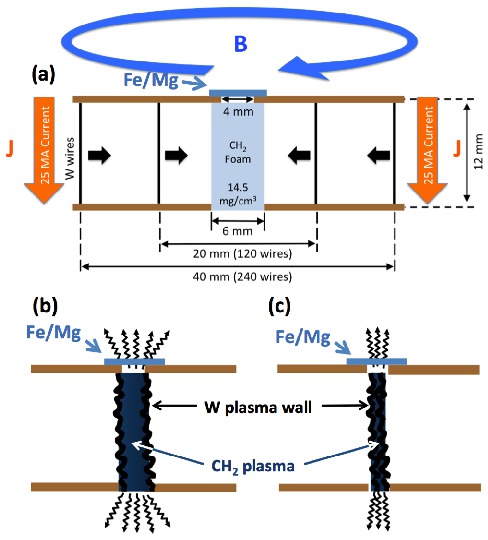}

\caption{\label{fig:how_ZPDH_works}Cross-sectional schematics of the cylindrical
z-pinch dynamic hohlraum (ZPDH). (a) Initial setup of the ZPDH, which
consists of 360 tungsten (W) wires concentrically arranged at two
radii with a cylindrical low-density plastic ($\mathrm{CH_{2}}$)
foam located at the z-pinch center. (b) When the imploding W plasma
collides with the $\mathrm{CH_{2}}$ foam, shock-generated radiation
is trapped inside the high-opacity W plasma wall and heats the sample.
(c) At stagnation, it provides an intense backlighter along the cylindrical
axis. }

\end{figure}

Fe opacity experiments were performed at the Sandia National Laboratories
Z-machine. The Fe sample is volumetrically heated by a z-pinch dynamic
hohlraum (ZPDH) and backlit at its stagnation. A detailed description
of the z-pinch dynamic hohlraum is discussed elsewhere\cite{Bailey:2006jz,Rochau:tf}.
Figure \ref{fig:how_ZPDH_works}(a) illustrates a cross-sectional
schematic of the cylindrical ZPDH initial setup. An azimuthal magnetic
field is generated by the electrical current, $J$, running through
tungsten wires strung in a cylindrical array. The current and self-generated
magnetic field produce a $\boldsymbol{J}\times\boldsymbol{B}$ force
that pushes the tungsten plasma towards the axis where a 14.5 $\mathrm{mg/cm^{3}}$
cylindrical plastic ($\mathrm{CH_{2}}$) foam is located. When the
tungsten plasma collides with the foam, it generates a radiative shock
in the $\mathrm{CH_{2}}$ {[}Fig. \ref{fig:how_ZPDH_works}(b){]}.
The shock radiation is trapped inside the thick tungsten plasma wall
due to the high opacity of the tungsten plasma. The Fe/Mg sample placed
on the top exit hole is heated by the hohlraum radiation, and it is
backlit when the radiating shock stagnates at the z-pinch axis {[}Fig.
\ref{fig:how_ZPDH_works}(c){]}.

\begin{figure}
\includegraphics[width=8.5cm]{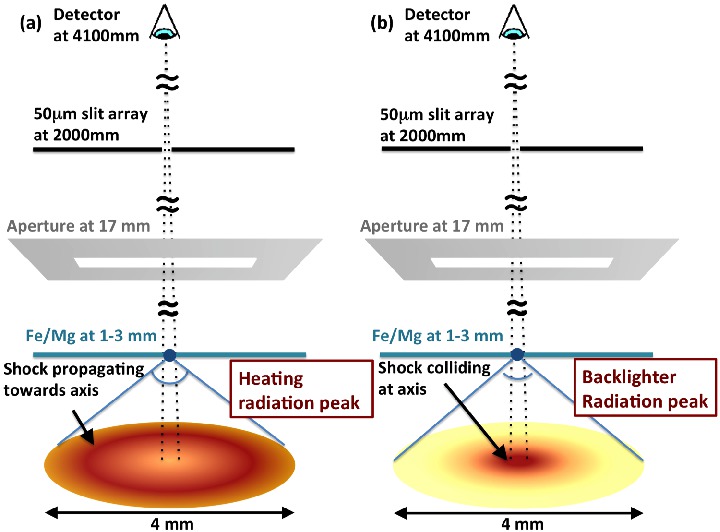}

\caption{\label{fig:ZPDH_radiation}Schematics showing the ZPDH radiation emitting
surfaces and the view factors from the samples and the detectors at
(a) the time of the heating radiation peak and (b) the time of the
backlighter radiation peak. Since each point on the sample sees the
entire emitting surface area, the sample experiences maximum radiation
(i.e., $\int I\cdot dA$) before stagnation. The detector records
most of the radiation during stagnation because its view is limited
to the central region by an aperture and 50 $\mu$m slits. The distances
specified for the various components are measured from the x-ray source. }
\end{figure}

Figure \ref{fig:ZPDH_radiation} illustrates the difference in the
ZPDH radiation seen by the sample and our spectrometers. While any
point in the Fe/Mg sample sees the entire ZPDH emitting surface, the
detector located $\sim$4100 mm away from the radiation source only
sees the central region of its surface because the detector view is
limited by an aperture and $\sim$50 $\mu$m slits, which are located
$\sim$17 mm and $\sim$2000 mm away from the source, respectively.
During a ZPDH implosion, the radiation intensity increases while the
area of the bright region decreases. The radiation power from the
ZPDH exit hole is proportional to intensity times emitting surface
area ($P=\int_{A}I\cdot dA$) and thus changes slowly compared to
the change in the implosion itself. The heating radiation is energetic:
its energy distribution maximum exceeds 700 eV. The Fe sample is almost
transparent to this radiation, and is therefore volumetrically heated\cite{Bailey:2009hh}.
As the implosion continues, the heating radiation power begins to
drop as the decrease in emitting area exceeds the increase in intensity.
A few nanoseconds after the heating radiation peak, the radiative
shock reaches the z-pinch axis, generating extremely bright radiation
that provides the backlighter for the transmission measurements {[}Fig.
\ref{fig:ZPDH_radiation}(b){]}. While the data are recorded by a
time-integrated spectrometer, time resolution is provided by the few-nanosecond
duration of the backlighter. 

The ZPDH backlighter is attenuated by the Fe/Mg sample, and the spectrum
is recorded with time-integrated potassium acid phthalate convex curved
crystal spectrometers equipped with Kodak RAR2492 x-ray film. The
spectrometers have between four and six slits $\sim$2000 mm away
from the sample and the backlighter, making the magnification $\sim$1.
The typical slit width is $\sim$50 $\mu$m, which provides a spatial
resolution of $\sim$100 $\mu$m. The crystal curvature radius is
4 or 6 inches, and an x-ray film is located at 8 cm from the crystal.
The spectral dispersion axis is determined by ray tracing using prominent
lines as wavelength references. Film optical densities are converted
to exposure based on Henke \textit{et al.} \cite{Henke:1984gp,Henke:1984ia}.
Filter transmissions and crystal reflectivity are corrected based
on the data available from the Center for X-Ray Optics with the instrument
geometry effects taken into account.

\begin{figure}
\includegraphics[width=8.5cm]{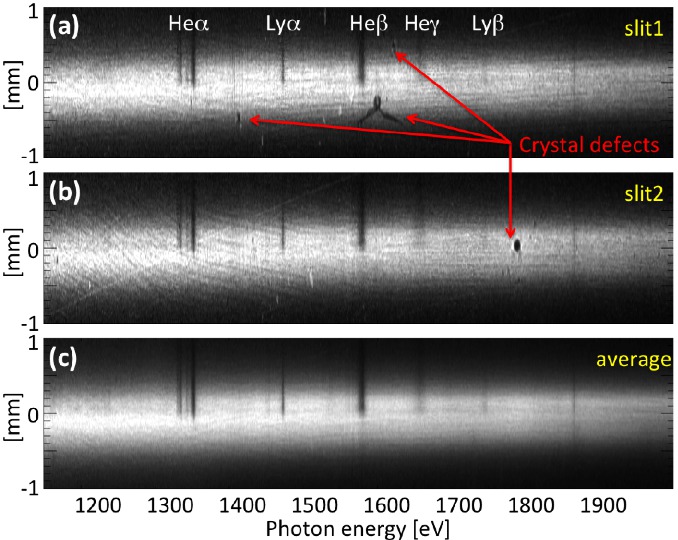}

\caption{\label{fig:data_reduction_average_image} Fe/Mg absorption images
over the Mg K-shell spectral range. Strong line features are from
Mg K-shell bound-bound line absorptions such as Mg He-$\alpha$, Ly-$\alpha$,
He-$\beta$, He-$\gamma$, and Ly-$\beta$. (a) and (b) show two single
slit images from z2364, which contain many crystal defects. (c) shows
the image averaged over six slit images of the same experiment. Averaging
improves the signal-to-noise ratio and helps to mitigate any bias
towards individual slit images due to crystal defects or other instrumental
problems. The spatial range of interest is from 100 $\mu m$ to 400
$\mu m$ where the backlighter is bright and the Fe/Mg exists in the
sample. In this experiment, there is no Fe/Mg in the bottom half of
the sample so that we have the option to extract transmission in a
single experiment by using the Fe/Mg attenuated spectrum from the
top of the image and the unattenuated spectrum from the bottom\cite{Bailey:2009hh}. }
\end{figure}

Recording multiple slit images has several advantages. First, one
can increase the signal-to-noise ratio of the data without giving
up the instrumental spatial resolution. Second, a single image could
be flawed by systematic issues such as crystal defects, slit width
variations along its length, or reflectivity variations over the crystal
surface. However, these systematic issues are random across the different
slit images and can be mitigated by averaging over multiple slit images.
For example, Fig. \ref{fig:data_reduction_average_image} (a) and
(b) show two of the six individual slit images of Fe/Mg absorption
measurements from a single experiment (z2364) and contain crystal
defects at random locations. When averaged over the six slit images
{[}Fig. \ref{fig:data_reduction_average_image} (c){]}, those artifacts
have been greatly reduced. 

\begin{figure}
\includegraphics[width=8.5cm]{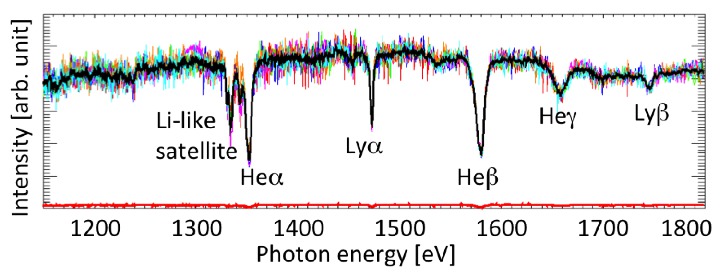}

\caption{\label{fig:Six-spectra-extracted} Fe/Mg absorption spectra from a
300 $\mu m$ wide region in Fig. \ref{fig:data_reduction_average_image}.
Prominent line features are Mg bound-bound line absorptions such as
Mg He-$\alpha$, Ly-$\alpha$, He-$\beta$, He-$\gamma$, and Ly-$\beta$.
Six spectra in different colors are extracted from individual slit
images. The averaged spectrum is overlaid in black, and its uncertainty
is shown at the bottom in red. The average percent uncertainty is
$\pm$2.5\%. }

\end{figure}

The averaged spectra can be extracted either from the averaged slit
image or by averaging the spectra extracted from individual slit images.
These two methods are identical as long as the crystal defects are
treated in the same way. The uncertainties in the spectra are estimated
by taking the standard deviation of the spectra extracted from N individual
slit images divided by $\sqrt{\mathrm{N}}$. An example is shown in
Fig. \ref{fig:Six-spectra-extracted}. The black and the red curves
are, respectively, the averaged spectrum and its uncertainty, which
for this data set is approximately $\pm2.5\%$.

\section{\label{sec:K-shell-spectroscopy-for}Plasma condition determination}

Fe conditions have to be measured independently in order to compare
the measured Fe transmission with the modeled transmission. Fe conditions
predicted by hydrodynamic simulations are not appropriate for this
purpose for two reasons. First, we are investigating Fe opacity because
the opacity model accuracy has been called into question. Since the
sample is heated by radiation in our experiments, the Fe opacity plays
a key role in the plasma evolution. If the Fe opacity is in question,
we should not expect the hyrdosimulation to predict plasma conditions
accurately. Second, hydrodynamic simulations require many input parameters.
Providing all of the required information for every experiment is
itself a challenge. Propagating all of the input uncertainties to
the Fe conditions uncertainties is even more challenging. Thus, we
measure Fe conditions directly through K-shell spectroscopy of Mg
mixed into the Fe sample.

While L-shell opacity and spectra are not completely understood, particularly
at high temperatures and densities, K-shell spectra from H-, He-,
and Li-like ions have been extensively researched and used to diagnose
plasma conditions\cite{Griem:1992jv,Hammel:dg,Bailey:2004bq,Bailey:2008cb}.
Due to the small number of bound electrons, the required atomic data
for singly and multiply excited states can be calculated with high
accuracy and fine structure detail. Additionally, many of the atomic
data have been experimentally validated\cite{NIST_ASD}. K-shell line
shapes for H-, He-, and Li-like ions have also been extensively investigated
over the last 50 years\cite{Baranger:1958gf,Smith:1966wn,griem1974spectral,Tighe:1978cga,Mancini:1991jr,Alexiou:2009jd,Iglesias:2010gh,Stambulchik:2010kj}
and are understood much better than L-shell line shapes. Different
K-shell spectral models would infer slightly different $T_{e}$ and
$n_{e}$ due to differences in details such as Stark line shape calculation,
continuum lowering, atomic data, and numerical approach. The Fe condition
uncertainties due to the spectral model details are under investigation
and will be discussed elsewhere.

\subsection{\label{sub:Temperature-and-density}Temperature and density sensitivity}

\begin{figure}
\includegraphics[width=8.5cm]{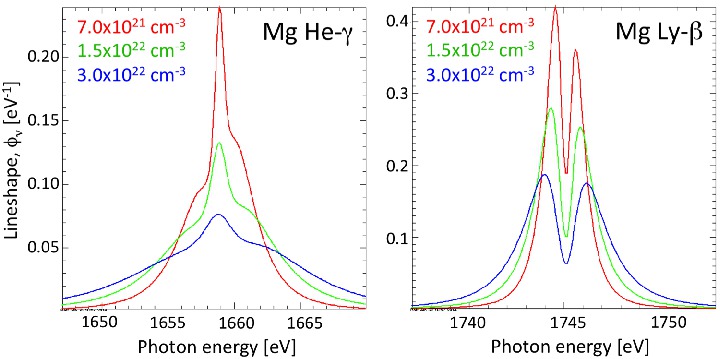}

\caption{\label{fig:density_from_stark_broadening} Area-normalized detailed
Stark line shapes calculated at $n_{e}$=$\mathrm{7.0\times10^{21}}$,
$\mathrm{1.5\times10^{22}}$, $\mathrm{3.0\times10^{22}\, cm^{-3}}$
for (a) Mg He-$\gamma$ and (b) Mg Ly-$\beta$. As density increases,
the line shapes become broader. }

\end{figure}

Line emission from K-shell ions is sensitive to temperature and density.
The bound-bound (bb) line transmission is defined as follows:
\begin{equation}
T_{\nu}^{bb}=\exp\left\{ -\tau_{\nu}^{bb}\right\} \label{eq:transmission}
\end{equation}
\begin{equation}
\tau_{\nu,\, lu}^{bb}=N_{Mg}\Delta x\cdot\frac{\pi e^{2}}{m_{e}c}f_{lu}\phi_{\nu,\, lu}p_{l}\label{eq:optical_depth}
\end{equation}
where $\tau_{\nu,\, lu}^{bb}$ is the line optical depth at frequency
$\nu$ due to photoexcitation from a lower level $l$ to an upper
level $u$, $N_{Mg}\Delta x$ is the Mg areal number density measured
by Rutherford backscattering, $f_{lu}$ is the oscillator strength
of the transition, $\phi_{\nu,\, lu}$ is the line shape associated
with the transition, and $p_{l}$ is the fractional population in
the lower state of the transition. When the oscillator strengths do
not depend on plasma conditions, the only unknowns are $\phi_{\nu,\, lu}$
and $p_{l}$. The line shape, $\phi_{\nu,\, lu}$, is determined by
Doppler broadening, natural broadening, and Stark broadening. We also
estimated the possible Zeeman effect on the line shape based on an
upper limit magnetic field in our experiments. The effect was confirmed
to be much smaller than the instrumental spectral resolution and thus
neglected in our line shape calculations. At the density of interest
here, the dominant line broadening mechanism is Stark broadening,
which is sensitive to electron density. Figure \ref{fig:density_from_stark_broadening}
shows area normalized line shapes for Mg He-$\gamma$ ($1s^{2}\rightarrow1s4p$)
and Mg Ly-$\beta$ ($1s\rightarrow3p$) calculated at three different
conditions using a detailed Stark line shape calculation code, MERL\cite{Mancini:1991jr}.
One can clearly see that, as electron density increases, the line
shapes become broader. This is more significant for the transitions
involving high principal quantum numbers, and thus, in this article,
the transitions with upper quantum number greater than 2 (e.g., $\beta$,
$\gamma$, $\delta$) are used to extract $n_{e}$. 

\begin{figure}
\includegraphics[width=8.5cm]{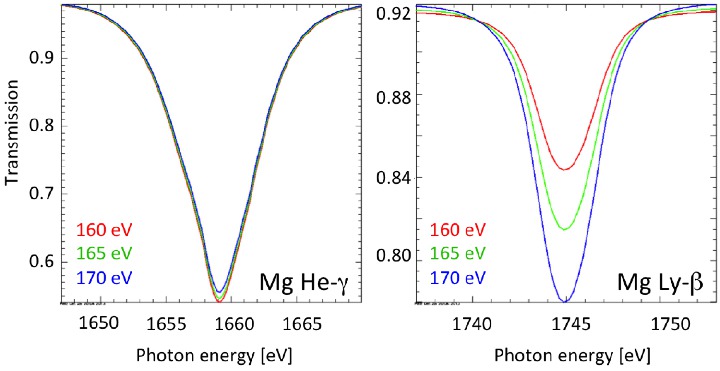}

\caption{\label{fig:Te_from_line_ratios}Mg He-$\gamma$ and Ly-$\beta$ transmissions
at $T_{e}$=160, 165, and 170 eV with a fixed electron density $n_{e}$=$8.0\times10^{21}\mathrm{\, cm^{-3}}$.
As temperature increases, the line ratio changes due to the change
in ionization balance. }

\end{figure}

To diagnose temperature, we use line ratios that reflect the $T_{e}$-dependent
ionization balance. The ratio of optical depths integrated over line
shapes with $\int\phi_{\nu}d\nu=1$ is:
\begin{eqnarray}
\frac{\tau_{Ly-\beta}^{bb}}{\tau_{He-\gamma}^{bb}} & \approx & \frac{f_{Ly-\beta}}{f_{He-\gamma}}\frac{p_{1s}}{p_{1s^{2}}}\nonumber \\
 & \propto & \frac{p_{1s}}{p_{1s^{2}}}\label{eq:line_ratio}
\end{eqnarray}
where $f_{Ly-\beta}$ and $f_{He-\gamma}$ are the oscillator strengths
of Ly-$\beta$ and He-$\gamma$ line transitions, respectively. Since
the H-like ground state population and He-like ground state population
are directly affected by charge state distributions, their ratio is
a strong temperature diagnostic at a given density. Figure \ref{fig:Te_from_line_ratios}
shows Mg He-$\gamma$ and Ly-$\beta$ line transmissions computed
at $T_{e}$=160, 165, and 170 eV with a fixed electron density $n_{e}$=$\mathrm{8.0\times10^{21}\, cm^{-3}}$
using a collisional radiative model, PrismSPECT in LTE mode\cite{MacFarlane:2003,MacFarlane:2007kj}.
As the temperature increases from 160 to 170 eV, the He-like ground
state population, $p_{1s^{2}}$, decreases by 4\% (0.831$\rightarrow$0.794)
while the H-like ground state population, $p_{1s}$, increases by
about a factor of two (0.051$\rightarrow$0.100), monotonically increasing
the ratio of these initial states, $p_{1s}/p_{1s^{2}}$, from $6.1\times10^{-2}$
to $\mathrm{12.6\times10^{-2}}$. Thus, from Eq. (\ref{eq:line_ratio}),
one can see that relative line strengths from adjacent charge states
are strongly dependent on the electron temperature.

Due to the monotonic relationships between line shapes and $n_{e}$
and between line ratios and $T_{e}$, the measured Mg lines constrain
the $T_{e}$ and $n_{e}$ of the Mg embedded region. Since Mg is mixed
throughout the Fe sample, the Fe plasma conditions can be inferred
from the Mg K-shell lines.

\subsection{Spectral model and parameter optimization}

In the present analysis, Mg spectra are computed by RADIATOR, which
is a framework to couple a tabulated emissivity and opacity database
with a radiation transport solver and compute emergent spectra by
taking into account a given instrumental broadening\cite{Nagayama:2012hh}.
This framework is combined with a multi-objective global optimization
program, GALM\cite{Nagayama:2012hh}, which is based on a genetic
algorithm (GA) followed by Levenberg-Marquardt non-linear least squares
minimization (LM)\cite{Goldberg:1989:GAS:534133,Press:1992:NRC:573140}.
The combination of the GA and LM is an efficient optimization algorithm
that finds the solution quickly by exploring small fractions of the
parameter space.

To use RADIATOR+GALM, two things have to be done: i) create a tabulated
emissivity and opacity for Mg and ii) add a radiation transport solver
for slab geometry to RADIATOR. The local emissivity and opacity database
for Mg is computed by PrismSPECT in LTE mode together with the detailed
Stark line shape database computed by MERL\cite{Mancini:1991jr}.
The details on how to extract local emissivity and opacity out of
the PrismSPECT calculation are discussed in Appendix A. A general
transport radiation solver is developed for slab geometry that can
take into account mixtures, linear gradients, background, and self-emission
options as needed. While the simple pure transmission approximation
is sufficient to extract the sample conditions, the general formalism
is useful to investigate various effects on the emergent spectra for
many applications and thus is discussed in detail in Appendix B. The
formalism is not exclusive to PrismSPECT and can be used with other
collisional radiative models\cite{Hansen:2007hu,2009PhRvE..80e6402F}.
The pure transmission approximation assumes that the backlighter is
much brighter than the plasma self-emission and is a slowly changing
function of frequency over the instrumental broadening width (Appendix
B):
\begin{equation}
T_{\nu}^{\star}=\int g\left(\nu-\nu'\right)T_{\nu'}d\nu'\label{eq:approximated_transmission}
\end{equation}
\begin{equation}
T_{\nu}=\exp\left\{ -N_{Mg}\Delta x\cdot\kappa_{\nu}^{PSDB,\, Mg}(T_{e},\, n_{e})\right\} \label{eq: transmission_PSDB}
\end{equation}
where the first equation convolves the computed transmission with
the instrumental broadening, $g\left(\nu-\nu'\right)$, and the second
equation is a transmission calculation using a tabulated Mg opacity
database computed by PrismSPECT (PSDB). $N_{Mg}\Delta x$ is the Mg
areal number density in $\mathrm{ions/cm^{-2}}$, and $\kappa_{\nu}^{PSDB,\, Mg}(T_{e},\, n_{e})$
is the tabulated frequency-dependent Mg fractional absorption coefficient
in $\mathrm{cm^{2}/ions}$. $\kappa_{\nu}^{PSDB,\, Mg}$ contains
bound-bound (bb), bound-free (bf), and free-free (ff) contributions:
\begin{equation}
\kappa_{\nu}^{PSDB,\, Mg}=\left(\kappa_{\nu}^{bb}+\kappa_{\nu}^{bf}+\kappa_{\nu}^{ff}\right)^{PSDB}\label{eq:PrismSPECT_opacity}
\end{equation}
which includes, 
\begin{equation}
\kappa_{\nu}^{bb}=\sum_{lu=all\, transitions}\left\{ \frac{\pi e^{2}}{mc}f_{lu}\phi_{\nu,\, lu}^{MERL}p_{l}\right\} \label{eq:b-b opacity}
\end{equation}
The MERL database used in PrismSPECT contains detailed line shapes
for Mg He-$\alpha$, $\beta$, $\gamma$, $\delta$, and $\epsilon$,
and Mg Ly-$\alpha$ and $\beta$. The ion microfields are computed
by APEX\cite{Iglesias:1985if} assuming an Fe:Mg mixture of 1:1. In
the temperature range of interest for this work, ion dynamic effects
are assumed to be negligible and omitted from the calculation. Both
temperature and density sensitivities are encoded in $\kappa_{\nu}^{PSDB,\, Mg}$
via the detailed line shape, $\phi_{\nu,\, lu}^{MERL}$, and the lower
population, $p_{l}$, in the $\kappa_{\nu}^{bb}$ contribution. The
areal number density $N_{Mg}\Delta x$ controls the line depths. RADIATOR+GALM
optimizes $T_{e}$, $n_{e}$, and $N_{Mg}\Delta x$ that best reproduce
the measured Mg spectra.

\subsection{Instrumental broadening effect on plasma condition diagnostics}

While the condition sensitivities exist in $p_{l}$ and $\phi_{\nu}$
of the optical depths $\tau_{\nu}$, what we measure is transmission
$T_{\nu}$ convolved with instrumental broadening. One can extract
$\tau_{\nu}$ out of $T_{\nu}$ perfectly if and only if the instrumental
spectral resolving power is infinite, which is never the case for
experiments. The spectral resolving power of our instrument was measured
to be $E/\Delta E$$\sim1000$\cite{Loisel:2012dv}. This finite spectral
resolution obscures $T_{e}$ and $n_{e}$ sensitivities embedded in
the transmission\cite{1990PhRvA..42.4788C}, and lines with higher
optical depths lose the sensitivities more significantly by this effect
due to the non-linear relationship between $\tau_{\nu}$ and $T_{\nu}$
{[}Eq. (\ref{eq:transmission}){]}. The rest of this section synthetically
illustrates how our instrumental spectral resolution affects the measured
lines, and determines what Mg lines are best used for the Fe plasma
$T_{e}$ and $n_{e}$ analysis. 

\begin{figure}
\includegraphics[width=8.5cm]{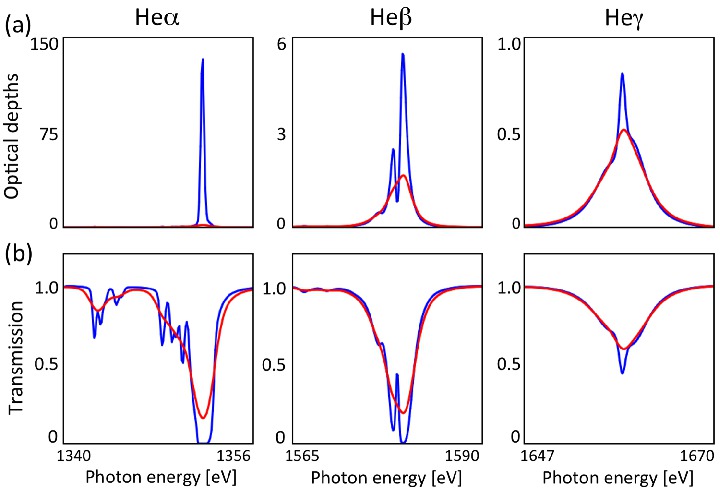}

\caption{\label{fig:weak_lines_should_be_fitted} (a) optical depths, $\tau_{\nu}$
and (b) transmission, $T_{\nu}$, of He-$\alpha$, $\beta$, $\gamma$
calculated at $T_{e}$=163 eV and $n_{e}$=$\mathrm{8\times10^{21}\, cm^{-3}}$
with $N_{Mg}\Delta x\approx7.4\times10^{17}\, ions/cm^{2}$. Blue
and red curves are before and after the instrumental broadening effects,
respectively. }

\end{figure}

The blue curves in Fig. \ref{fig:weak_lines_should_be_fitted} (a)
and (b) are the calculated optical depths, $\tau_{\nu}$, and transmissions,
$T_{\nu}$, respectively, for Mg He-$\alpha$ ($1s^{2}-1s2p$), He-$\beta$
($1s^{2}-1s3p$), and He-$\gamma$ ($1s^{2}-1s4p$) at $T_{e}$=163
eV, $n_{e}$=$\mathrm{8\times10^{21}\, cm^{-3}}$, and $\mathrm{N_{Mg}\Delta x}\approx7.4\times10^{17}\,\mathrm{ion/cm^{2}}$.
These are typical conditions achieved in our experiments. While these
lines share the same initial state population $p_{1s^{2}}$ in Eq.
(\ref{eq:optical_depth}), their optical depths are very different
due to the differences in their oscillator strengths and line shapes.
When the modeled transmissions are convolved with the instrumental
spectral shape, transmissions at the line centers are overestimated,
and all the detailed structures are smoothed out as shown in the red
curves in Fig. \ref{fig:weak_lines_should_be_fitted}(b). One then
cannot successfully recover the calculated optical depths when converting
from these convolved transmissions, reducing our ability to accurately
diagnose $T_{e}$ and $n_{e}$. Red lines in Fig. \ref{fig:weak_lines_should_be_fitted}(a)
show optical depths converted from transmission with the instrumental
spectral resolution effect. One can clearly see that the saturation
due to the instrumental broadening affects the stronger lines more.
While the He-$\gamma$ line preserves both the shape and the strength,
the He-$\alpha$ and $\beta$ lines are heavily altered. The saturation
in the He-$\alpha$ is particularly severe and the red line is barely
visible in Fig. 7(a). When optical depths are larger than 1, the apparent
line shape and strength are strongly affected by instrumental broadening.
Thus, in this article, we decided to analyze weaker Mg lines ($\tau_{\nu}\lesssim$1)
for the purpose of the iron plasma $T_{e}$ and $n_{e}$ diagnostics.

\subsection{Mg bound-bound \textit{line} transmission}

\begin{figure}
\includegraphics[width=8.5cm]{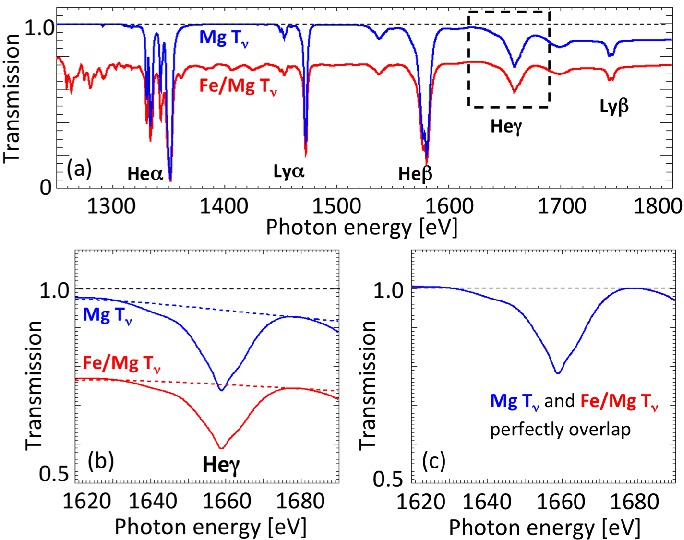}

\caption{\label{fig:b-b transmission}This figure compares calculated transmission
spectra from an Fe/Mg mixture (red) and from pure Mg (blue). (a) Fe/Mg
transmission is lower than Mg transmission due to additional attenuation
by Fe b-f but does not have much Fe b-b contribution except $h\nu<$1450
eV. (b) blow-up of the dashed region of (a). The red and blue dashed
lines are baselines defined by replacing the apparent Mg b-b lines
by straight lines. (c) Mg b-b line transmission spectra defined by
dividing transmission spectra {[}solid curves in (b){]} by the baselines
{[}dashed lines in (b){]}. }

\end{figure}

Transmission spectra can be extracted by dividing Fe/Mg absorption
spectra by unattenuated backlighter spectra. However, fitting the
modeled Mg transmission spectra directly to the measured transmission
spectra is difficult because the measured transmission has additional
Fe attenuation in the Mg b-b line spectral region. Figure \ref{fig:b-b transmission}(a)
shows Fe/Mg mixture (red) and pure Mg (blue) synthetic transmission
spectra computed at the same conditions. Because of the Fe attenuation,
Fe/Mg mixture transmission is lower than pure Mg transmission. Also,
while Fe attenuation in this spectral range is smooth and slowly varying
with photon energy, there are smooth b-f contributions from Mg above
1700 eV. It is difficult to objectively separate Mg b-f from the Fe
b-f in the measured data. Fitting the modeled Fe/Mg transmission spectra
to the measured transmission spectra is not ideal because we do not
want to rely on calculated Fe opacity to diagnose the Fe conditions.
Since plasma condition sensitivity mostly comes from Mg b-b lines
as in Eq. \ref{eq:transmission}, one strategy is to extract Mg b-b
line transmission spectra both from the measured Fe/Mg spectra and
the modeled Mg transmission spectra, and then compare them. 

To this end, we first remove Mg b-b lines and then replace them with
straight lines to define baselines. By dividing the spectra by the
baselines, one can extract Mg b-b line transmission spectra. This
removes not only Fe b-f but also Mg b-f from the spectra. For the
modeled Mg spectra, we could compute the b-b line transmission for
specific lines of interest as in Eq. \ref{eq:transmission}. However,
the details of composite spectral formation such as overlapping lines,
satellite line contributions, and continuum lowering could affect
the instrumental broadening, the baseline determination, and the emergent
spectral line shapes. Thus, complete transmission spectra are computed
first, and then the baselines are defined to extract b-b line transmission
spectra in the same way as for the data. Figures \ref{fig:b-b transmission}(b)
and (c) illustrate how baselines are defined for the synthetic data
in Fig. \ref{fig:b-b transmission}(a) and how the resultant Mg b-b
line transmission spectra agree between Fe/Mg and Mg spectra. The
red and blue dashed lines in Fig. \ref{fig:b-b transmission}(b) are
the baselines defined for Fe/Mg and pure Mg spectra, respectively.
By dividing the spectra by the baselines, Mg b-b line transmission
spectra are extracted {[}Fig. \ref{fig:b-b transmission}(c){]}. This
study shows that Mg b-b spectra extracted from Fe/Mg spectra are identical
to those from pure Mg spectra. Thus, Fe conditions can be inferred
by fitting modeled spectra to the measured spectra in Mg b-b line
transmissions.

There are two comments on Mg b-b line transmission analysis. One is
on the Fe/Mg mixture effects on the emergent spectra. In this article,
we assume LTE. In LTE plasmas, level populations are fully determined
by $T_{e}$ and $n_{e}$ regardless of whether the plasma is a mixture
or not, and mixture effects will only appear in the line shapes via
ion microfields contributed from both Fe and Mg ions. As long as mixture
effects are included in the Mg Stark line shapes calculation, Mg b-b
lines extracted from Fe/Mg spectra and from pure Mg spectra should
be identical {[}Fig. \ref{fig:b-b transmission}(c){]}. Another comment
is on condition uncertainty due to the baseline determination from
the measured spectra. This is a concern because there is noise in
the measured spectra. However, baselines are determined based on many
data points on both sides of b-b lines, and thus the baseline uncertainty
is smaller than the uncertainty of any of those data points. The uncertainty
due to the baseline determination is included in the uncertainty due
to experiment-to-experiment variation in the following sections. 

\section{\label{sec:Fe-condition-analysis}Measured Fe plasma conditions}

In order to disentangle the complex physical processes included in
opacity models and benchmark those models, it is important to measure
Fe opacity at different conditions, but with similar charge state
distributions. Nash \textit{et al.} used LASNEX 2D\cite{1978JOSA...68..549Z}
hydrodynamic simulations to explore a way to control Fe sample conditions
by changing the target configuration\cite{Nash:2010cn}. They predicted
that the Fe sample $n_{e}$ could be controlled by changing the rear
$\mathrm{CH}$ tamper thickness. Adding more tamping mass on the back
slows the expansion speed and maintains higher density at the time
of the backlight. The slower expansion would also produce modestly
higher temperatures.

\begin{figure}
\includegraphics[width=8.5cm]{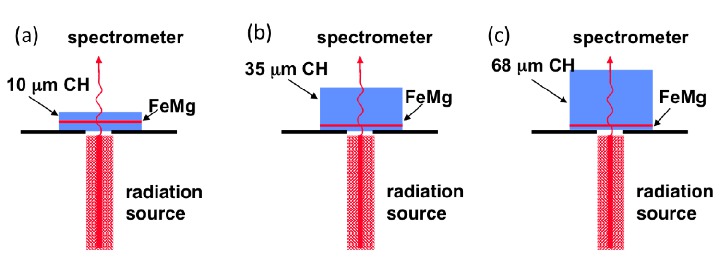}

\caption{\label{fig:Target_designs}Fe/Mg samples are tamped by three different
thicknesses of plastic ($\mathrm{CH}$). (a) rear: 10 $\mu$m, front:
10 $\mu$m (b) rear: 35 $\mu$m, front: 2 $\mu$m, (c) rear: 68 $\mu$m,
front: 2 $\mu$m. Based on the hydrosimulation, thicker rear $\mathrm{CH}$
tamper would reach higher $T_{e}$ and $n_{e}$ in Fe/Mg plasmas.
The pictures are not to scale, but exaggerated to illustrate the different
tamper thicknesses. }

\end{figure}

Based on their suggestions, Fe opacity experiments were performed
with three different rear $\mathrm{CH}$ tamper thicknesses as shown
in Fig. \ref{fig:Target_designs}: (a) two experiments with 10 $\mu$m,
(b) one experiment with 35 $\mu$m, and (c) six experiments with 68
$\mu$m. By increasing this rear $\mathrm{CH}$ tamper thickness,
the backlighter radiation is attenuated more. In order to minimize
this extra attenuation, we reduced the front $\mathrm{CH}$ thickness
from 10 $\mu$m to 2 $\mu$m for (b) and (c) assuming that Fe/Mg samples
do not expand downward due to the pressure provided by the ZPDH $\mathrm{CH_{2}}$
foam plasma.

\begin{figure}
\includegraphics[width=8.5cm]{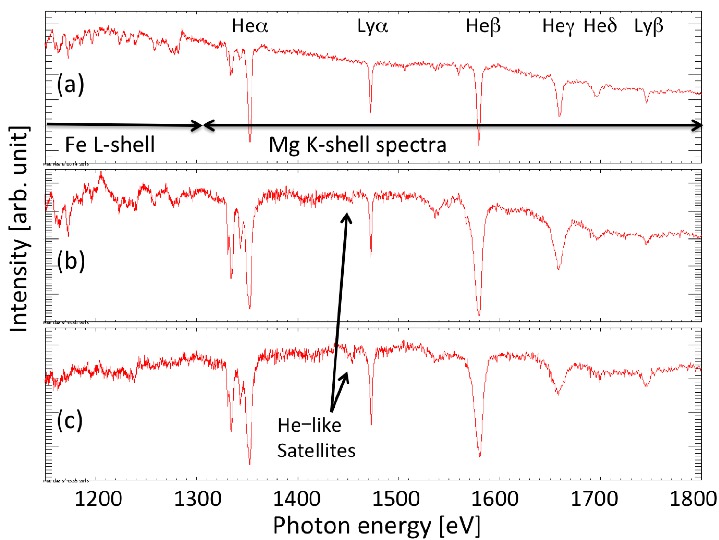}

\caption{\label{fig: whole_Fe_Mg_spectra} Fe/Mg spectra recorded through the
samples with rear $\mathrm{CH}$ tamper thicknesses of (a) 10 $\mu$m,
(b) 35 $\mu$m, (c) 68 $\mu$m.}
\end{figure}

In this section, we summarize the Fe sample conditions for each target
configuration, which are analyzed by Mg K-shell line transmission
spectroscopy. Figure \ref{fig: whole_Fe_Mg_spectra} (a), (b), and
(c) show the Fe/Mg absorption spectra recorded from the different
target configurations shown in Fig. \ref{fig:Target_designs} (a),
(b), and (c), respectively. One can qualitatively observe that Mg
K-shell lines become broader as the rear $\mathrm{CH}$ tamper thickness
increases, which indicates that $n_{e}$ becomes higher with increasing
tamper thickness. For example, He-$\delta$ becomes broader from Fig.
\ref{fig: whole_Fe_Mg_spectra}(a) to (b), and becomes even broader
and almost merged into the continuum in (c). Also, He-like satellite
lines of Ly-$\alpha$ are not visible in (a), become visible in (b),
and even more prominent in (c). These transitions are $1s2p\rightarrow2p^{2}$
or $1s2s\rightarrow2s2p$ and start from excited levels. This is proof
that, as going from (a) to (c), there is a larger fraction of the
population in excited levels and a sign that both $T_{e}$ and $n_{e}$
are higher. This means that the plasma must be hot and dense enough
for the collisional excitation rates to be comparable to the spontaneous
radiative decay rates. The actual condition of each sample is quantitatively
analyzed based on the method discussed in Sec. \ref{sec:K-shell-spectroscopy-for}.

\subsection{10 $\mu$m rear $\mathrm{CH}$ tamper}

\begin{figure}
\includegraphics[width=8.5cm]{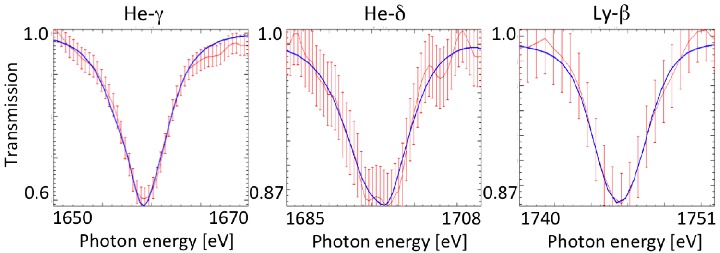}

\caption{\label{fig:thin_CH_results}Model fits (blue) to the Mg He-$\gamma$,
He-$\delta$, and Ly-$\beta$ lines recorded from experiment 2301
(red). The inferred conditions are $T_{e}=166\pm2$ eV and $n_{e}=\left(6.3\pm0.4\right)\times10^{21}\mathrm{cm^{-3}}$.}

\end{figure}

First, we analyzed Fe conditions using the thin $\mathrm{CH}$ tamper
target {[}Fig. \ref{fig:Target_designs} (a){]}, where the Fe/Mg sample
is sandwiched by 10 $\mathrm{\mu m}$ thick plastic. Figure \ref{fig: whole_Fe_Mg_spectra}
(a) shows the Fe/Mg absorption spectrum from z2301 (atomic ratio of
Fe:Mg=0.5:1). The Mg lines with $\tau_{\nu}<1$ are He-$\gamma$,
He-$\delta$, and Ly-$\beta$. Optical depths for Ly-$\alpha$ and
He-$\beta$ are approximately 4 and 6, respectively, and excluded
from the analysis. Thus, the analysis focuses on He-$\gamma$, He-$\delta$,
and Ly-$\beta$, which are simultaneously analyzed with RADIATOR+GALM.
The inferred conditions are $T_{e}=166\pm2$ eV and $n_{e}=\left(6.3\pm0.4\right)\times10^{21}\mathrm{cm^{-3}}$.
The individual fits are shown in Fig. \ref{fig:thin_CH_results}.
Another experiment, z2221 , used the same $\mathrm{CH}$ configuration
and Fe:Mg=0.9:1. The simultaneous analysis of its Mg He-$\gamma$,
He-$\delta$, and Ly-$\beta$ infers $T_{e}=167\pm4$ eV and $n_{e}=\left(7.9\pm1.5\right)\times10^{21}\,\mathrm{cm^{-3}}$,
in agreement with z2301 within uncertainties. The means and the standard
deviations from these two experiments are $\left\langle T_{e}\right\rangle =167\pm1$
eV and $\left\langle n_{e}\right\rangle =\mathrm{\left(7.1\pm1.1\right)\times10^{21}\, cm^{-3}}$
where these standard deviations indicate the experiment-to-experiment
variation in the Fe conditions. The mean measurement uncertainties
are $\left(2+4\right)/2=3$ eV and $\mathrm{\left(0.4+1.5\right)\times10^{21}/2=1.0\times10^{21}\, cm^{-3}}$.
Total uncertainties are computed by adding the mean measurement uncertainties
to the experimental variation in quadrature: $\sqrt{1^{2}+3^{2}}\approx3$
eV and $\mathrm{\sqrt{1.1^{2}+1.0^{2}}\times10^{21}\approx1.5\times10^{21}\, cm^{-3}}$.
Thus, the Fe conditions inferred from the thin rear $\mathrm{CH}$
tamper target are $T_{e}=\mathrm{167\pm3}$ eV and $n_{e}=\mathrm{\left(7.1\pm1.5\right)\times10^{21}\, cm^{-3}}$.
This density result is consistent with the density reported in 2007;
however, this temperature is significantly higher than the corresponding
temperature, $156\pm6$ eV \cite{Bailey:2007fq,Bailey:2008cb}. One
reason is that the data reported in 2007 were recorded before the
Z-machine refurbishment of that same year, which increased the electrical
power delivered to the load. This difference is evidence that the
refurbished Z-machine produces higher radiation power, therefore reaching
higher temperatures in the Fe sample.

\subsection{35 $\mu$m rear $\mathrm{CH}$ tamper}

\begin{figure}
\includegraphics[width=8.5cm]{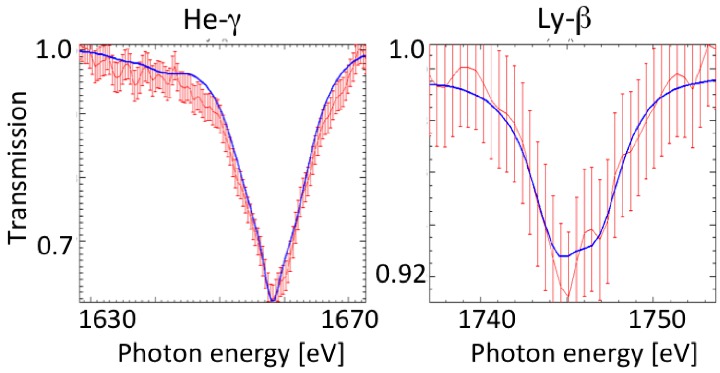}

\caption{\label{fig:Fits_intermed_CH}Mg He-$\gamma$ and Ly-$\beta$ are simultaneously
fit for experiment z2363. The best fit is provided by $T_{e}=170$
eV and $n_{e}=2.0\times10^{22}\,\mathrm{cm^{-3}}$. }
\end{figure}

We performed only one experiment, z2363, using the 35 $\mu$m rear
$\mathrm{CH}$ tamper target, whose spectra are shown in Fig. \ref{fig: whole_Fe_Mg_spectra}(b).
The Fe:Mg atomic ratio is 0.9:1. The lines with $\tau_{\nu}<1$ are
Mg He-$\gamma$ and Ly-$\beta$. These lines are simultaneously analyzed,
and the fits are given in Fig. \ref{fig:Fits_intermed_CH}(a) and
(b), respectively. The inferred conditions are $T_{e}=170\pm2$ eV
and $n_{e}=\left(2.0\pm0.2\right)\times10^{22}\,\mathrm{cm^{-3}}$.
We confirmed a slight increase in $T_{e}$ and more than a factor
of two increase in $n_{e}$ compared to the conditions achieved by
the 10 $\mu$m rear $\mathrm{CH}$ tamper target. Since there is only
one experiment from this target configuration, the uncertainties do
not include experiment-to-experiment sample conditions reproducibility. 

\subsection{\label{sub:68-m-thick}68 $\mu$m thick rear $\mathrm{CH}$ tamper}

\begin{figure}
\includegraphics[width=8.5cm]{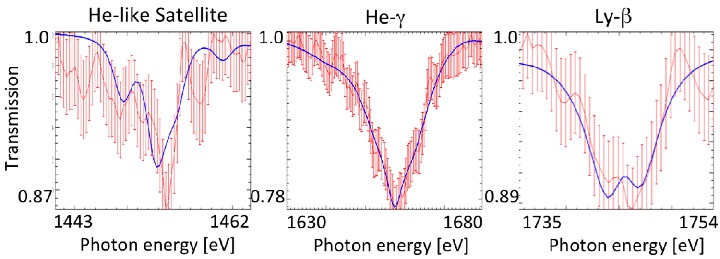}

\caption{\label{fig:thick_CH_results}Fits to Mg He-like satellites, He-$\gamma$,
and Ly-$\beta$ recorded on experiment z2364. The inferred conditions
are $T_{e}=195\pm3$ eV and $n_{e}=\left(4.3\pm0.5\right)\times10^{22}\,\mathrm{cm^{-3}}$.}

\end{figure}

We performed six experiments using the 68 $\mu$m rear $\mathrm{CH}$
tamped target illustrated in Fig. \ref{fig:Target_designs}(c). Mg
lines with $\tau_{\nu}<$1 are He-$\gamma$, Ly-$\beta$, and the
He-like satellite lines of Ly-$\alpha$. Thus, these lines from z2364
{[}i.e., Fig. \ref{fig: whole_Fe_Mg_spectra}(c){]} are simultaneously
analyzed using RADIATOR+GALM. The fits to these lines are shown in
the Fig. \ref{fig:thick_CH_results}, and the inferred conditions
are $T_{e}=195\pm3$ eV and $n_{e}=\left(4.3\pm0.5\right)\times10^{22}\,\mathrm{cm^{-3}}$. 

\begin{figure}
\includegraphics[width=8.5cm]{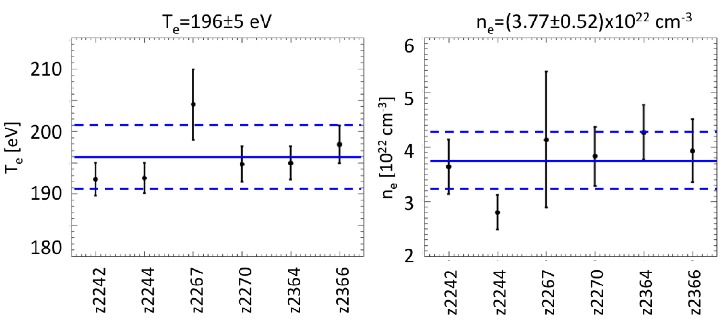}

\caption{\label{fig:thick_CH_all_results}Fe condition analysis result summary
for thick rear $\mathrm{CH}$ tamper targets. The average and standard
deviation are indicated by blue solid and dashed lines, respectively:
$T_{e}=196\pm5$ eV and $n_{e}=\left(3.8\pm0.5\right)\times10^{22}\,\mathrm{cm^{-3}}$.
The experiment-to-experiment variations of $T_{e}$ and $n_{e}$ over
the six experiments are $\pm$3\% and $\pm$13\%, respectively. }

\end{figure}

While the $\mathrm{CH}$ configurations are the same for the six experiments,
Fe areal number densities are different to test the reliability of
the measured Fe opacity by applying Beer's law\cite{Bailey:2009hh}.
Figure \ref{fig:thick_CH_all_results} shows the conditions inferred
for each of the six experiments in (a) $T_{e}$ and (b) $n_{e}$.
Experiments z2364 and z2366 were performed with the thinnest Fe samples,
whose areal densities were $\sim\mathrm{0.66\times10^{18}\, Fe/cm^{2}}$
(atomic ratio of Fe:Mg=0.5:1). Experiments z2242 and z2270 were performed
with intermediate Fe thicknesses with areal number densities of $\sim1.2\times10^{18}\,\mathrm{Fe/cm^{2}}$
(Fe:Mg=1:1). Experiments z2244 and z2267 were performed with the thickest
Fe samples with areal number densities of $\sim\mathrm{2.9\times10^{18}\, Fe/cm^{2}}$
(Fe:Mg=2.3:1). We did not observe any correlation between Fe thickness
and the inferred Fe conditions. The average and standard deviation
are indicated by blue solid and dashed lines, respectively: $T_{e}=196\pm5$
eV and $n_{e}=\left(3.8\pm0.5\right)\times10^{22}\,\mathrm{cm^{-3}}$.
Based on the standard deviations, the Fe condition reproducibility
is $\pm$3\% in $T_{e}$ and $\pm$13\% in $n_{e}$. The mean individual
measurement uncertainties are 3 eV and 0.6$\times\mathrm{10^{22}\, cm^{-3}}$.
The total $T_{e}$ and $n_{e}$ uncertainties are computed by adding
the experiment-to-experiment variations and the mean individual measurement
uncertainties in quadrature, which are 6 eV ($\pm$3\%) and $\mathrm{0.8\times10^{22}\, cm^{-3}}$
($\pm$21\%), respectively. 

\section{\label{sec:Uniformity-measurement}Uniformity measurement}

\begin{figure}
\includegraphics[width=8.5cm]{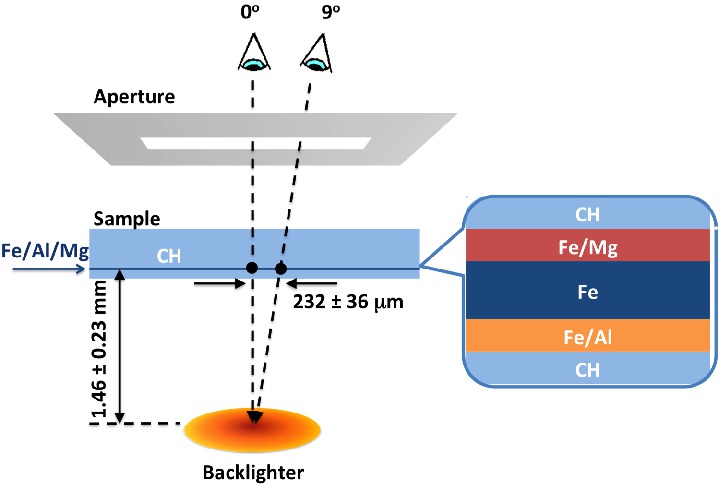}

\caption{\label{fig:uniformity_measurement_design}Spectrometers on axis and
at 9$^{\circ}$ sample transmission spectra from slightly different
spatial centers. The analyses of the measured spectra would reveal
not only axial non-uniformity but also a lateral gradient if it were
significant. }
\end{figure}

Uniformity of the Fe sample is asserted based on volumetric heating
provided by the powerful ZPDH radiation. In Sec. \ref{sec:Fe-condition-analysis},
similar Fe conditions are inferred from different Fe thicknesses,
which also supports the sample axial uniformity. However, since volumetric
heating is never perfect and heating radiation is supplied from one
side, it is worthwhile to examine the axial uniformity assumption
with more explicit experimental evidence. To this end, an experiment
was designed to investigate the sample axial non-uniformity. Instead
of mixing Mg throughout the Fe sample, Mg is mixed in the observer
(rear) side of Fe and another dopant, Al (Z=13), is mixed in the radiation
source (front) side of the Fe (Fig. \ref{fig:uniformity_measurement_design}),
which are separated by a pure Fe region. One can infer Fe conditions
in the radiation source side and in the observer side by Al and Mg
K-shell spectroscopy, respectively.

\begin{figure}
\includegraphics[width=8.5cm]{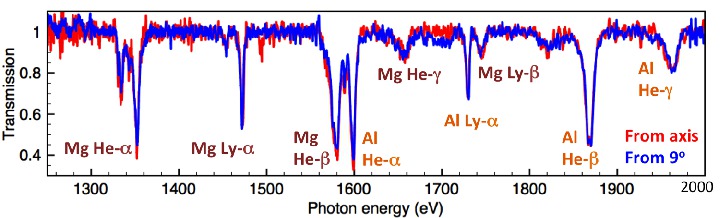}

\caption{\label{fig:Al_Mg_spectra}Comparison of line transmission spectra
recorded along the vertical axis (at 0$^{\circ}$) and at 9$^{\circ}$.
The good agreement indicates lateral gradients are insignificant.}
\end{figure}

The resultant Fe/Al/Mg (atomic ratio of 2:1:1) absorption spectra
were recorded along two different lines of sight; one along the vertical
axis (at 0$^{\circ}$) and the other at 9$^{\circ}$ from the vertical
axis as in Fig. \ref{fig:uniformity_measurement_design}. The distance
between the sample and the backlighter is measured on similar experiments,
based on parallax, and found to be 1.5$\pm$0.2 mm. Due to this source-to-sample
distance, when the spectrometers observe the backlighter through the
Fe/Al/Mg sample, they see through slightly different spatial regions
in the sample, which are separated by 232$\pm$36 $\mu$m as indicated
in Fig. \ref{fig:uniformity_measurement_design}. The spectra are
extracted by integrating the backlighter images over the central 300
$\mu$m. Since their backlighter centers appear at different points
in the sample, which are separated by 236 $\mu$m, 300 $\mu$m integrations
span different lateral regions in the sample with a small amount of
overlap ($\sim$64 $\mu$m). Thus, in addition to measuring an axial
gradient, these data will reveal a lateral gradient if it is significant.
Figure \ref{fig:Al_Mg_spectra} shows the extracted line transmission
spectra. In addition to Mg, lines from Al are apparent. 

\begin{figure}
\includegraphics[width=8.5cm]{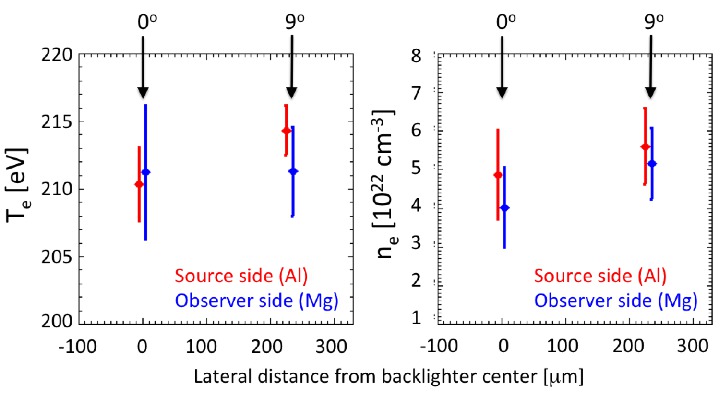}

\caption{\label{fig:uniformity_results}Measured $T_{e}$ (left) and $n_{e}$
(right) from uniformity analysis. The agreement between the conditions
measured in the radiation source side (red) and observer side (blue)
of the sample confirm that there is no axial gradient within uncertainties.
Similarly, the agreement between the 0$^{\circ}$ and $9^{\circ}$
line of sight conditions show that there is no measurable lateral
gradient.}

\end{figure}

The lines with $\tau_{\nu}<1$ are He-$\gamma$ and Ly-$\beta$ from
Mg and Ly-$\alpha$ and He-$\gamma$ from Al. These spectra are analyzed
using RADIATOR+GALM and the inferred $T_{e}$ and $n_{e}$ are summarized
in Table \ref{tab:Sample-uniformity-analysis}. Figure \ref{fig:uniformity_results}
shows the resultant $T_{e}$ (left) and $n_{e}$ (right). Red and
blue data points are the plasma conditions inferred for the radiation
source side (Al mixed side) and the observer side (Mg mixed side)
of the Fe sample, respectively. The red and the blue points agree
within their measurement uncertainties. Thus, it is confirmed that
there is no measurable axial gradient in the sample. Also, the conditions
inferred from different spectrometers (i.e., 0$^{\circ}$ and $9^{\circ}$)
agree within their uncertainties. These results indicate that there
is no gradient over $\pm230\,\mu$m from the heating center (i.e.,
the point in the sample where the backlighter center appears from
the spectrometer at $0^{\circ}$).

\begin{table}
\begin{centering}
\begin{tabular}{|c|c|c|}
\hline 
$0^{\circ}$ & $T_{e}$ {[}eV{]} & $n_{e}$ {[}$\mathrm{10^{22}\, cm^{-3}}${]}\tabularnewline
\hline 
\hline 
Al side & $\mathrm{210\pm3}$ & $\mathrm{4.9\pm1.2}$\tabularnewline
\hline 
Mg side & $\mathrm{211\pm5}$ & $\mathrm{4.0\pm1.0}$\tabularnewline
\hline 
\end{tabular}
\par\end{centering}

\begin{centering}
\begin{tabular}{|c|c|c|}
\hline 
$9^{\circ}$ & $T_{e}$ {[}eV{]} & $n_{e}$ {[}$\mathrm{10^{22}\, cm^{-3}}${]}\tabularnewline
\hline 
\hline 
Al side & $\mathrm{214\pm2}$ & $\mathrm{5.6\pm1.0}$\tabularnewline
\hline 
Mg side & $\mathrm{211\pm3}$ & $\mathrm{5.2\pm0.9}$\tabularnewline
\hline 
\end{tabular}
\par\end{centering}

\caption{\label{tab:Sample-uniformity-analysis}Sample uniformity analysis
results. No measurable axial nor lateral gradients are observed. }

\end{table}

\section{Conclusions}

We addressed three of the key opacity measurement requirements: sample
conditions control, independent sample condition measurements, and
sample uniformity verification. The opacity experiments were performed
at the Sandia National Laboratories Z-machine to benchmark opacity
models for Fe at the base of solar convection zone where $T_{e}$
and $n_{e}$ are 185 eV and $\mathrm{9\times10^{22}\, cm^{-3}}$,
respectively. The Fe sample conditions were controlled by the tamping
$\mathrm{CH}$ thicknesses. Increasing the rear $\mathrm{CH}$ thickness
slows down the expansion, and the Fe samples remain dense at the time
of backlighting. Three different rear $\mathrm{CH}$ thicknesses are
tested: i) 10 $\mu$m, ii) 35 $\mu$m, and iii) 68 $\mu$m. The resultant
Fe conditions are independently measured by Mg K-shell spectroscopy
and found to be i) $T_{e}$=167$\pm3$ eV and $n_{e}=\mathrm{\left(7.1\pm1.5\right)\times10^{21}\, cm^{-3}}$,
ii) $T_{e}$=170$\pm2$ eV and $n_{e}=\mathrm{\left(2.0\pm0.2\right)\times10^{22}\, cm^{-3}}$,
and iii) $T_{e}$=196$\pm6$ eV and $n_{e}=\mathrm{\left(3.8\pm0.8\right)\times10^{22}\, cm^{-3}}$,
respectively. One experiment was designed and performed to specifically
investigate the sample uniformity, with Al mixed into the side of
the sample facing the radiation source (front), and Mg mixed in the
opposite side of the sample (rear). The conditions at the bottom and
the top of the Fe sample were inferred by Al and Mg K-shell spectroscopy,
respectively, and confirmed that there are no measurable gradients
in the sample. The inferred condition uncertainties due to K-shell
spectral model details are under investigation, as well as the effects
of non-LTE, sample self-emission, and temporal gradients. While different
K-shell spectral models might infer slightly different $T_{e}$ and
$n_{e}$, those differences should be systematic and would not affect
the conclusions on the condition reproducibility and the sample uniformity.
The sample condition uncertainties due to K-shell spectral model details
are under investigation.

Fe opacity experiments at three different conditions with similar
charge state distributions provide crucial information to disentangle
complex physical processes in opacity models. Opacity models have
to accurately compute relevant atomic data, atomic level populations,
and then spectral formation, taking into account correct line shapes,
and the complexity is enhanced as temperature and density increase.
In LTE, the same average ionizations achieved by different $T_{e}$
and $n_{e}$ have different population distribution within each charge
state because of the temperature dependence on the Boltzmann relationship
{[}$p_{excited}/p_{ground}\propto\exp\left(-\Delta E/kT\right)${]}.
As temperature increases, more population is in excited states and
absorption starting from those excited states becomes more important.
Thus, at higher temperature, the accuracy and the treatment of excited
states (i.e., singly-, doubly-, ..., multiply-excited states) become
more crucial to accurately solve for population and model opacity.
As density increases, the effects of continuum lowering and Stark
broadening become more important. Since our data are measured at different
$T_{e}$ and $n_{e}$ with similar charge states, we can study how
these effects gradually change the true opacity, and how well state-of-the-art
opacity models can calculate them.

\section*{Acknowledgement }

We are grateful to R. Falcon and T. Lockard for their help in refining
the manuscript. Sandia is a multiprogram laboratory operated by Sandia
Corporation, a Lockheed Martin Company, for the United States Department
of Energy under contract DE-AC04-94AL85000.

\section*{Appendix A: Extraction of local emissivity and opacity from PrismSPECT}

PrismSPECT is a collisional radiative model that computes atomic level
populations and then computes opacity (i.e., mass absorption coefficient),
$\kappa_{\nu}^{mass}$, and emergent spectral radiance, $I_{\nu}$\cite{MacFarlane:2003,MacFarlane:2007kj}.
In slab geometry, these two outputs are related by the following equations:
\begin{equation}
I_{\nu}=\frac{\epsilon_{\nu}}{\kappa_{\nu}}\left[1-\exp\left(-\kappa_{\nu}L\right)\right]\label{eq:radiation_transport_slab}
\end{equation}
\begin{equation}
\kappa_{\nu}=\kappa_{\nu}^{mass}\rho\label{eq:absorption_coefficient}
\end{equation}
where $L$ is the thickness of the slab, and $\epsilon_{\nu}$ and
$\kappa_{\nu}$ are emission coefficient (or $emissivity$) and absorption
coefficient (or $opacity$), respectively\cite{mihalas1978stellar}.
For the rest of Appendix A and B, opacity refers to absorption coefficient,
$\kappa_{\nu}$ (not mass absorption coefficient, $\kappa_{\nu}^{mass}$).
Since both emissivity and opacity are proportional to the ion number
density, $n_{ion}$, it is convenient to invert the PrismSPECT outputs,
$I_{\nu}$ and $\kappa_{\nu}^{mass}$, into what we call fractional
emissivity and opacity as follows: 
\begin{equation}
\epsilon_{\nu}^{frac}\equiv\frac{\epsilon_{\nu}}{n_{ion}}=\frac{m_{ion}I_{\nu}\kappa_{\nu}^{mass}}{1-\exp\left(-\kappa_{\nu}^{mass}\rho L\right)}\label{eq:fractional_emissivity}
\end{equation}
\begin{equation}
\kappa_{\nu}^{frac}\equiv\frac{\kappa_{\nu}}{n_{ion}}=m_{ion}\kappa_{\nu}^{mass}\label{eq:fractional_opacity}
\end{equation}
The quantities $\epsilon_{\nu}^{frac}$ and $\kappa_{\nu}^{frac}$
are proportional to the initial state fractional population of relevant
radiative processes and depend only on $T_{e}$ and $n_{e}$ in the
LTE assumption. The database of $\epsilon_{\nu}^{frac}\left(T_{e},\, n_{e}\right)$
and $\kappa_{\nu}^{frac}\left(T_{e},\, n_{e}\right)$ can be very
useful for quick spectra calculations. One can build a PrismSPECT
fractional emissivity and opacity database of an element $X$, $\epsilon_{\nu}^{frac,\, X}\left(T_{e},\, n_{e}\right)$
and $\kappa_{\nu}^{frac,\, X}\left(T_{e},\, n_{e}\right)$, by performing
a single element, LTE PrismSPECT calculation for ranges of temperature
and density and extracting $\epsilon_{\nu}^{frac}$ and $\kappa_{\nu}^{frac}$
from each condition using Eqs. (\ref{eq:fractional_opacity}) and
(\ref{eq:fractional_emissivity}). 

We note that PrismSPECT requires ion number density, $n_{ion}$, as
an input (not $n_{e}$). Thus, $n_{e}$ has to be extracted from the
PrismSPECT output using $n_{e}=Z^{*}n_{ion}$, where $Z^{*}$ is the
mean charge of the plasma. Interpolations are required both to build
the database and to use the database. The required database grid spacing
in $T_{e}$, $n_{e}$, and $h\nu$, and PrismSPECT calculation details
depend on the required accuracy, and have to be carefully investigated
for each application. The database used in this article is optimized
for the emergent spectra accuracy, and temperature and density diagnostics.
The spectra based on the database with linear interpolation agree
with the spectra directly computed by PrismSPECT within 1\%. The uncertainties
in inferred $T_{e}$ and $n_{e}$ due to the use of the database and
the linear interpolation are within 0.5 \% and 1\%, respectively.
For Mg and Al database calculations, PrismSPECT employed tabulated
Stark line shapes for their K-shell lines, which were computed in
detail by MERL\cite{Mancini:1991jr}. 

The advantage of the database is speed and flexibility. Fractional
emissivity and opacity of an element are dependent only on $T_{e}$
and $n_{e}$ and independent of surrounding species, its own ion number
density, and geometry. Thus, once the database is extracted for each
element, one can compute spectra for either single or multiple elements,
uniform or non-uniform, emission or absorption by solving radiation
transport for a given geometry, without re-running collisional radiative
models. The database extraction is not exclusive to PrismSPECT; databases
can be extracted from other collisional models. The accuracy of the
database depends on the details of the collisional model used and
the numerical details of the extraction scheme. The idea of the database
can be extended for non-LTE as long as the non-LTE effects are independent
of geometry such as for the optically-thin approximation or for non-LTE
effects dominated by strong external radiation.

\section*{Appendix B: Slab radiation transport solver for a mixture}

Once fractional emissivity and opacity databases are computed for
each element $X$ {[}i.e., $\epsilon_{\nu}^{frac,\, X}\left(T_{e},\, n_{e}\right)$
and $\kappa_{\nu}^{frac,\, X}\left(T_{e},\, n_{e}\right)$ in Appendix
A{]}, one can solve the slab radiation transport equation to compute
emergent spectra for any mixture and any gradient. Geometry can even
be extended to arbitrary shapes\cite{Nagayama:2012hh}. Assume there
is an M species plasma, which is discretized into N zones along the
observation line of sight where zone N is closest to the observer.
By assuming that an event is observed very far from the plasma, emergent
spectra seen from the observer can be computed by the following equation:
\begin{equation}
I_{\nu}=I_{\nu,0}\exp(-\tau_{\nu})+I_{\nu,N}^{plasma}+F_{\nu}\label{eq:emergent_spectra}
\end{equation}
where $I_{\nu,\,0}$ is the incident spectral irradiance (or backlighter),
$\tau_{\nu}$ is the net optical depth of the plasma, $I_{\nu,\, N}^{plasma}$
is the plasma self-emission taking into account its self-absorption
effect, and $F_{\nu}$ is a potential additional background. $I_{\nu,\, N}^{plasma}$
can be computed recursively as follows:

\begin{equation}
\mathrm{I_{\nu,\, i}^{plasma}=I_{\nu,\, i-1}^{plasma}\exp\left(-\tau_{\nu,\, i}\right)+\frac{\epsilon_{\nu,\, i}}{\kappa_{\nu,\, i}}\left[1-\exp\left(-\tau_{\nu,\, i}\right)\right]}\label{eq:radiation_transport_mixture3-1}
\end{equation}
\begin{equation}
I_{\nu,\,1}^{plasma}=\frac{\epsilon_{\nu,\,1}}{\kappa_{\nu,\,1}}\left[1-\exp\left(-\tau_{\nu,\,1}\right)\right]\label{eq:self_emission}
\end{equation}
 
\begin{equation}
\epsilon_{\nu,\, i}=\sum_{j=1}^{M}N_{i}^{j}\epsilon_{\nu}^{frac,\, j}\left(T_{e,\, i},\, n_{e,\, i}\right)\label{eq:em}
\end{equation}

\begin{equation}
\kappa_{\nu,\, i}=\sum_{j=1}^{M}N_{i}^{j}\kappa_{\nu}^{frac,\, j}\left(T_{e,\, i},\, n_{e,\, i}\right)\label{eq:op}
\end{equation}
\begin{equation}
\eta_{\nu,\, i}=\epsilon_{\nu,\, i}L_{i}\text{ and }\tau_{\nu,\, i}=\kappa_{\nu,\, i}L_{i}\label{eq:emop_x_L}
\end{equation}
where $I_{\nu,\, i}^{plasma}$ is the emergent spectral irradiance
at the end of zone $i$ of the discretized plasma, and $\epsilon_{\nu,\, i}$
and $\kappa_{\nu,\, i}$ are local emissivity and opacity (i.e., absorption
coefficient) in zone $i$, respectively. The local emissivity and
opacity at zone $i$ (i.e., $\epsilon_{\nu,\, i}$ and $\kappa_{\nu,\, i}$)
can be computed by summing element emissivity, $N_{i}^{j}\epsilon_{\nu}^{frac,\, j}\left(T_{e,\, i},\, n_{e,\, i}\right)$,
and opacity, $N_{i}^{j}\kappa_{\nu}^{frac,\, j}\left(T_{e,\, i},\, n_{e,\, i}\right)$,
over all of the species in zone $i$, where $N_{i}^{j}$ are the ion
number density of element $j$ at zone $i$, and $T_{e,\, i}$ and
$n_{e,\, i}$ are electron temperature and density at zone $i$, respectively.
By multiplying $\epsilon_{\nu,\, i}$ and $\kappa_{\nu,\, i}$ by
the zone length, $L_{i}$, one can compute self-emission in the optically-thin
approximation, $\eta_{\nu,\, i}$, and the optical depth, $\tau_{\nu,\, i}$,
of zone $i$, respectively. Furthermore, $\eta_{\nu}=\sum_{i}\eta_{\nu,\, i}$
and $\tau_{\nu}=\sum_{i}\tau_{\nu,\, i}$ gives the net self-emission
in optically-thin approximation and the net optical depth of the plasma,
respectively. 

This formula is general and works for either emission or absorption
spectra. If the incident flux (or backlighter) is weaker than the
sample self emission (i.e., $I_{\nu,\,0}<I_{\nu,\, N}^{plasma}$),
Eq. (\ref{eq:radiation_transport_mixture3-1}) gives emission spectra.
If the backlighter is brighter (i.e., $I_{\nu,\,0}>I_{\nu,\, N}^{plasma}$),
this formula naturally produces absorption spectra. The measured spectra
can be simulated by convolving $I_{\nu}$ with instrument spectral
shape, $g\left(\nu-\nu'\right)$, as follows:
\[
I_{\nu}^{\star}=\int g\left(\nu-\nu'\right)I_{\nu'}d\nu'
\]
where $\int g(\nu-\nu')d\nu'=1$. The measured transmission spectra,
$T_{\nu}^{\star}$, can be computed by dividing $I_{\nu}^{\star}$
by the convolved backlighter, $\int g\left(\nu-\nu'\right)I_{\nu',\,0}d\nu'$. 

There are two limiting cases of Eq. (\ref{eq:radiation_transport_mixture3-1}).
For these limiting cases, $F_{\nu}$ is neglected for communication
purposes. One limiting case is the optically-thin approximation for
emission spectra (i.e., $\kappa_{\nu,\, i}=0$ and $B_{\nu}=0$).
This approximation simplifies Eq. (\ref{eq:radiation_transport_mixture3-1})
to $I_{\nu}=\eta_{\nu}$. The other limiting case is pure transmission
or the bright-backlighter approximation (i.e., $I_{\nu,\,0}\gg I_{\nu,\, N}^{plasma}$),
which simplifies Eq. (\ref{eq:radiation_transport_mixture3-1}) to
the following:
\[
I_{\nu}=I_{\nu,\,0}\exp(-\tau_{\nu})
\]
When the backlighter is not only bright but also spectrally smooth,
one can approximate the instrumental broadening effect as follows:
\begin{eqnarray*}
T_{\nu}^{\star} & = & \frac{\int g\left(\nu-\nu'\right)I_{\nu'}d\nu'}{\int g\left(\nu-\nu'\right)I_{\nu',\,0}d\nu'}\\
 & \approx & \frac{\int g\left(\nu-\nu'\right)I_{\nu'}d\nu'}{I_{\nu,\,0}\int g\left(\nu-\nu'\right)d\nu'}\\
 & \approx & \int g\left(\nu-\nu'\right)\frac{I_{\nu'}}{I_{\nu,\,0}}d\nu'\\
 & \approx & \int g\left(\nu-\nu'\right)T_{\nu'}d\nu'
\end{eqnarray*}
which is Eq. (\ref{eq:approximated_transmission}). This approximation
is valid when $I_{\nu,\,0}$ changes slowly over the instrumental
spectral shape, $g\left(\nu-\nu'\right)$. 

In our application, areal number densities of each element (i.e.,
Fe, Mg, and Al) are measured prior to the experiments using Rutherford
backscattering (RBS). They can also be inferred by RADIATOR+GALM because
of their sensitivities to the line depths. Thus, rewriting the $\epsilon_{\nu,\, i}$/$\kappa_{\nu,\, i}$,
$\eta_{\nu,\, i}$, and $\tau_{\nu,\, i}$ in terms of element areal
number density, $\left(N^{j}L\right)$, and fraction of the element
areal number density in each zone $i$, $\alpha_{i}^{j}=\left(N_{i}^{j}L_{i}\right)/\left(N^{j}L\right)$,
is also useful:
\[
\frac{\epsilon_{\nu,\, i}}{\kappa_{\nu,\, i}}=\frac{\epsilon_{\nu,\, i}L_{i}}{\kappa_{\nu,\, i}L_{i}}=\frac{\eta_{\nu,\, i}}{\tau_{\nu,\, i}}
\]
\[
\eta_{\nu,\, i}=\sum_{j}\epsilon_{\nu,\, i}^{frac,\, j}\alpha_{i}^{j}\left(N^{j}L\right)
\]
\[
\tau_{\nu,\, i}=\sum_{j}\kappa_{\nu,\, i}^{frac,\, j}\alpha_{i}^{j}\left(N^{j}L\right)
\]
where $\sum_{i}\alpha_{i}^{j}=1.0$. By assuming the fraction $\alpha_{i}^{j}$
is determined by the initial target design\cite{Nagayama:2012hs},
one can compute emergent spectra of multi-species plasmas just by
providing $T_{e,\, i}$ and $n_{e,\, i}$ at each zone $i$. The areal
number density of each element, $\left(N^{j}L\right)$, can either
be provided by RBS measurements or extracted from the analysis.

%

\end{document}